\documentclass[a4paper,11pt]{article}

\usepackage{amsfonts,amssymb,amsmath,amsthm,dsfont,xfrac,xspace}
\usepackage{fullpage}
\usepackage{graphicx}
\usepackage{subfig}
\usepackage{float}
\usepackage{cite}
\usepackage{hyperref}
\usepackage{algorithmic}
\graphicspath{{./},{fig/}}
\makeatletter
\let\NAT@parse\undefined
\makeatother
\usepackage[sort&compress, numbers]{natbib}
\usepackage{hyperref}

\usepackage{mathtools,bbm}
\usepackage{cuted}
\usepackage{multicol}
\usepackage{stfloats}

\newsavebox{\ieeealgbox}

\renewcommand{\thelem}{\the \numexpr (\value{thm}+1) \relax.\arabic{lem}}

\title{Comments on ``Probabilities of Error for Adaptive Reception of \emph{M}-Phase Signals''} 

\author{Trung-Hien~Nguyen\thanks{T.-H.~Nguyen,  P.~De~Doncker, and F.~Horlin are with OPERA department, Universit\'e libre de Bruxelles (ULB), 1050 Brussels, Belgium. E-mail: trung-hien.nguyen@ulb.ac.be} , J\'er\^ome~Louveaux\thanks{J.~Louveaux is with ICTEAM institute, Universit\'e catholique de Louvain (UCL), 1348 Louvain-la-Neuve, Belgium.} , Philippe~De~Doncker$^*$, and Fran\c cois~Horlin$^*$  } 

\begin{document}
\maketitle

\begin{abstract}
	Paper [1] derived the probability density function (PDF) of a sum of products of two correlated complex Gaussian zero-mean random variables (RVs) that has been applied to calculate the error probabilities of a \emph{M}-ary phase shift keying (\emph{M}-PSK) system. We show that there exist some typos in the computation and we provide the detailed derivation leading to the correct expressions.
\end{abstract}


\section{Introduction}

The paper~\cite{ProakisTCT68} has received a large interest for many years as it is an important milestone in the analytical derivation of error probabilities in the communication system based on \emph{M}-ary phase shift keying (\emph{M}-PSK) modulations. However, there exist some typos in the equations \cite[eq.~(11)]{ProakisTCT68} and \cite[eq.~(12)]{ProakisTCT68} deriving the probability density function (PDF) of the dot-product between the complex-valued vectors of signal and of channel, referred to as the random variable (RV) $Z$. These typos will be corrected in this letter. Observing closely the goals of the paper~\cite{ProakisTCT68}, we note two major points:

\begin{itemize}
\item The correctness of the PDF of $Z$ is important, as it affects directly the correctness of subsequent derived PDF of the phase (or amplitude) of $Z$ that is required for the error probability calculation. Unfortunately, the detailed derivation was omitted in~\cite{ProakisTCT68} and only the final formula is given.
\item The PDF should be a function of the variances of signal RV and channel RV as well as of the correlation between these two RVs. However, when the correlation is set to $0$ (i.e., no correlation), it appears that the joint PDFs~\cite[eqs. (11),~(12)]{ProakisTCT68} are independent of the variances, which suggest us that the expressions are erroneous.
\end{itemize}

\noindent Therefore, the aims of this letter are three-fold:

\begin{itemize}
\item The computation of the PDF of $Z$.
\item The exact derivation of PDF of the amplitude and of the phase of $Z$. In addition to the solution proposed in~\cite{ProakisTCT68}, a new alternative approximation of the PDF of the phase of $Z$ is derived, which contains elementary trigonometric functions that are interesting for the further error probability calculation of \emph{M}-PSK systems.
\item The numerical verification of the correctness of derived PDFs.
\end{itemize}

\emph{Notation}: ${\mathbb{E}}[\cdot]$ and $| \cdot |$ are the expectation and absolute operators, respectively; $X_R$ and $X_I$ denote the real and imaginary parts of $X$, respectively; $x!$ is the factorial of a positive integer $x$; $\Gamma (\cdot)$ is the Gamma function and $\Gamma(x) = (x-1)!$ for a positive integer number $x$; $\mathbb{J}_L (\cdot)$ is the $L$-th order Bessel function of the first kind; $\mathbb{I}_L (\cdot)$ and $\mathbb{K}_L (\cdot)$ are the $L$-th order modified Bessel functions of the first and second kind, respectively.

\section{Corrected derivation of [1, (11), (12)]}

We first remind the problem mentioned in~\cite{ProakisTCT68}: considering that $( X_l, Y_l )$ are complex-valued, zero-mean, Gaussian RVs, i.e., ${X_l} \sim \mathcal{CN} \left( {0,\sigma _X^2} \right)$, ${Y_l} \sim \mathcal{CN} \left( {0,\sigma _Y^2} \right)$ of cross-correlation $\mu  = {{{\rm{\mathbb{E}}}\left[ {{X_l} \cdot {Y_l}} \right]} \mathord{\left/ {\vphantom {{{\rm{\mathbb{E}}}\left[ {{X_l} \cdot {Y_l}} \right]} {\sqrt {{\rm{\mathbb{E}}}\left[ {{{\left| {{X_l}} \right|}^2}} \right] \cdot {\rm{\mathbb{E}}}\left[ {{{\left| {{Y_l}} \right|}^2}} \right]} }}} \right. \kern-\nulldelimiterspace} {\sqrt {{\rm{\mathbb{E}}}\left[ {{{\left| {{X_l}} \right|}^2}} \right] \cdot {\rm{\mathbb{E}}}\left[ {{{\left| {{Y_l}} \right|}^2}} \right]} }} = \left| \mu  \right| \cdot {{\rm{e}}^{j\varepsilon }}$; and such that $( X_l, Y_l )$ are statistically independent and identically distributed with any other pairs ($X_k, Y_k$) for $\forall l \ne k$, we need to derive the PDF of $Z = \sum\nolimits_{l = 1}^L {{X_l} \cdot {Y_l}} $.

In order to derive the PDF of $Z$, we follow the steps suggested in~\cite{ProakisTCT68} as follows:

\begin{itemize}
\item Step 1: Derive the joint characteristic function (CF) of $Z_R$ and $Z_I$.
\item Step 2: Perform the inverse CF transformation to yield the joint PDF of $Z_R$ and $Z_I$.
\item Step 3: Perform the Cartesian-polar transformation to yield the joint PDF of the amplitude $R = \sqrt {Z_R^2 + Z_I^2} $ and the phase $\Theta  = {\arctan }\left( {{{{Z_I}} \mathord{\left/ {\vphantom {{{Z_I}} {{Z_R}}}} \right. \kern-\nulldelimiterspace} {{Z_R}}}} \right)$.
\item Step 4: Integrate the joint PDF either over the RV $\Theta$ to yield the PDF of $R$ or over the RV $R$ to obtain the PDF of $\Theta$.
\end{itemize}

\subsection{Step 1}

We first consider the case $L=1$ and omit the subscripts $l$ of RVs $X_l$ and $Y_l$ for simplicity. Once we find the CF of $Z$, the generalization with any value of $L$ is straightforward thanks to the properties of the CF. Let us express $Y$ as the contribution of $X$ plus a RV $U \sim \mathcal{CN} \left( {0,\sigma _U^2} \right)$ independent of $X$ as follows:

\begin{equation}
\begin{dcases}
{Y_R} & = \frac{{{\sigma _Y}\left| \mu  \right|\cos \varepsilon }}{{{\sigma _X}}}{X_R} + \frac{{{\sigma _Y}\left| \mu  \right|\sin \varepsilon }}{{{\sigma _X}}}{X_I} + {U_R} \\
{Y_I}   & = \frac{{{\sigma _Y}\left| \mu  \right|\sin \varepsilon }}{{{\sigma _X}}}{X_R} - \frac{{{\sigma _Y}\left| \mu  \right|\cos \varepsilon }}{{{\sigma _X}}}{X_I} + {U_I}
\end{dcases}
\label{eq1}
\end{equation}

\noindent where the variance $\sigma _{{U_R}}^2 = \sigma _{{U_I}}^2 = \frac{{\sigma _U^2}}{2} = \frac{{\sigma _Y^2}}{2}\left( {1 - {{\left| \mu  \right|}^2}} \right)$. It is easy to check that the correlation between $X$ and $Y$ is $\mu$. From aforementioned assumptions and the distributions of $X$ and $Y$, it is straightforward to obtain the distribution of $Z_R$ and $Z_I$ conditioned on $X$ as follows:

\begin{equation}
{\left. {{Z_R}} \right| X} \sim \mathcal{N} \left( {\frac{{{\sigma _Y}\left| \mu  \right|\cos \varepsilon }}{{{\sigma _X}}}{{\left| X \right|}^2},\frac{{\sigma _Y^2}}{2}\left( {1 - {{\left| \mu  \right|}^2}} \right){{\left| X \right|}^2}} \right) 
\label{eq2}
\end{equation}

\begin{equation}
{\left. {{Z_I}} \right| X} \sim \mathcal{N} \left( {\frac{{{\sigma _Y}\left| \mu  \right|\sin \varepsilon }}{{{\sigma _X}}}{{\left| X \right|}^2},\frac{{\sigma _Y^2}}{2}\left( {1 - {{\left| \mu  \right|}^2}} \right){{\left| X \right|}^2}} \right) 
\label{eq3}
\end{equation}

The joint CF of $Z_R$ and $Z_I$ conditioned on $X$ can be expressed by:

\begin{align}
{\Psi _{\left. {{Z_R},{Z_I}} \right|X}}(j{\omega _1},  {\rm{  }}{\left. {j{\omega _2}} \right| X})   &  = {\mathbb{E}}\left[ {{{\left. {\exp \left( {j\left( {{\omega _1}{z_R} + {\omega _2}{z_I}} \right)} \right)} \right|} {X = x}}} \right]  \nonumber  \\
{\rm{    }} &  = \exp \left\{ {j\frac{{{\sigma _Y}\left| \mu  \right|}}{{{\sigma _X}}}\left( {{\omega _1}\cos \varepsilon  + {\omega _2}\sin \varepsilon } \right){{\left| x \right|}^2}} \right.      \left. { - \frac{{\sigma _Y^2}}{4}\left( {1 - {{\left| \mu  \right|}^2}} \right)\left( {\omega _1^2 + \omega _2^2} \right){{\left| x \right|}^2}} \right\}  .
\label{eq4}
\end{align}

The joint CF of $Z_R$ and $Z_I$ can now be derived as:

\begin{equation}
{\Psi _{{Z_R},{Z_I}}}\left( {j{\omega _1},j{\omega _2}} \right) = \int\limits_{ - \infty }^\infty  {{\Psi _{\left. {{Z_R},{Z_I}} \right|X}}\left( {{{\left. {j{\omega _1},j{\omega _2}} \right|} X}} \right){f_X}\left( x \right)dx} .
\label{eq6}
\end{equation}

\noindent where the PDF of the RV $X$ is expressed by~\cite{ProakisBook00} ${f_X}\left( x \right) = \frac{1}{{\pi \sigma _X^2}}{ {\rm{exp}} { \left( - \frac{{{{\left| x \right|}^2}}}{{\sigma _X^2}} \right) } }$.

Substituting \eqref{eq4} into \eqref{eq6} and changing variables, i.e., ${x_R} = t\cos \phi $ and ${x_I} = t\sin \phi $, after some mathematical manipulations, \eqref{eq6} can be rewritten as follows:

\begin{align}
{\Psi _{{Z_R},{Z_I}}}\left( {j{\omega _1},j{\omega _2}} \right)   =    &  \  \frac{1}{{\pi \sigma _X^2}}\int\limits_0^\infty  {\int\limits_0^{2\pi } {t \cdot \exp \left\{ { - \frac{{{t^2}}}{{\sigma _X^2}}} \right.} }     + j\frac{{{\sigma _Y}\left| \mu  \right|}}{{{\sigma _X}}}\left( {{\omega _1}\cos \varepsilon  + {\omega _2}\sin \varepsilon } \right){t^2}   \nonumber  \\
{\rm{                              }}  &  \left. { - \frac{{\sigma _Y^2}}{4}\left( {1 - {{\left| \mu  \right|}^2}} \right)\left( {\omega _1^2 + \omega _2^2} \right){t^2}} \right\}d\phi \ dt
\label{eq7}
\end{align}

By solving the trivial problem \eqref{eq7}, the CF corresponding to $L=1$ is obtained. The CF is generalized to any value of $L$ as presented in \eqref{eq8}, in which we apply the property that the CF of the summation of independent RVs is equal to the multiplication of all individual CFs w.r.t each RV. It can be observed that we obtain again the CF derivation as in~\cite{ProakisTCT68}.

\begin{figure*}[!htbp]
    \begin{equation}
	  {\Psi _{{Z_R},{Z_I}}}\left( {j{\omega _1},j{\omega _2}} \right) = {\left[ {\frac{{\frac{4}{{\sigma _X^2\sigma _Y^2\left( {1 - {{\left| \mu  \right|}^2}} \right)}}}}{{{{\left( {{\omega _1} - j\frac{{2\left| \mu  \right|\cos \varepsilon }}{{{\sigma _X}{\sigma _Y}\left( {1 - {{\left| \mu  \right|}^2}} \right)}}} \right)}^2} + {{\left( {{\omega _2} - j\frac{{2\left| \mu  \right|\sin \varepsilon }}{{{\sigma _X}{\sigma _Y}\left( {1 - {{\left| \mu  \right|}^2}} \right)}}} \right)}^2} + \frac{4}{{\sigma _X^2\sigma _Y^2{{\left( {1 - {{\left| \mu  \right|}^2}} \right)}^2}}}}}} \right]^L}
       \label{eq8}
    \end{equation}
\end{figure*}

\subsection{Step 2} 

The joint PDF of $Z_R$ and $Z_I$ can be derived by performing the inverse CF transformation as follows:

\begin{align}
{f_{{Z_R},{Z_I}}}\left( {{z_R},{z_I}} \right) = \frac{1}{{4{\pi ^2}}}\int\limits_{ - \infty }^\infty        &    {\int\limits_{ - \infty }^\infty  {{\Psi _{{Z_R},{Z_I}}}\left( {j{\omega _1},j{\omega _2}} \right)} }    \cdot {{\rm{e}}^{ - j\left( {{\omega _1}{z_R} + {\omega _2}{z_I}} \right)}}d{\omega _1}d{\omega _2} .
\label{eq9}
\end{align}

By changing the variables ${t_1} = {\omega _1} - j\frac{{2\left| \mu  \right|\cos \varepsilon }}{{{\sigma _X}{\sigma _Y}\left( {1 - {{\left| \mu  \right|}^2}} \right)}}$ and ${t_2} = {\omega _2} - j\frac{{2\left| \mu  \right|\sin \varepsilon }}{{{\sigma _X}{\sigma _Y}\left( {1 - {{\left| \mu  \right|}^2}} \right)}}$, \eqref{eq9} can be rewritten as:

\begin{align}
{f_{{Z_R},{Z_I}}}\left( {{z_R},{z_I}} \right) =  &  \frac{{{4^{L - 1}}{{\left( {1 - {{\left| \mu  \right|}^2}} \right)}^{ - L}}}}{{{\pi ^2}{{\left( {{\sigma _X}{\sigma _Y}} \right)}^{2L}}}} \cdot {{\rm{e}}^{\frac{{2\left| \mu  \right|\left( {{z_R}\cos \varepsilon  + {z_I}\sin \varepsilon } \right)}}{{{\sigma _X}{\sigma _Y}\left( {1 - {{\left| \mu  \right|}^2}} \right)}}}}       \int\limits_{ - \infty }^\infty  {\int\limits_{ - \infty }^\infty  {{{\left( {\frac{4}{{\sigma _X^2\sigma _Y^2{{\left( {1 - {{\left| \mu  \right|}^2}} \right)}^2}}} + t_1^2 + t_2^2} \right)}^{ - L}}} }    \nonumber  \\
{\rm{                                  }} &  \qquad \quad  \cdot {{\rm{e}}^{ - j\left( {{z_R}{t_1} + {z_I}{t_2}} \right)}}d{t_1}d{t_2} \ .
\label{eq10}
\end{align}

Let $A = \frac{{{4^{L - 1}}{{\left( {1 - {{\left| \mu  \right|}^2}} \right)}^{ - L}}}}{{{\pi ^2}{{\left( {{\sigma _X}{\sigma _Y}} \right)}^{2L}}}} \cdot \exp \left( {\frac{{2\left| \mu  \right|\left( {{z_R}\cos \varepsilon  + {z_I}\sin \varepsilon } \right)}}{{{\sigma _X}{\sigma _Y}\left( {1 - {{\left| \mu  \right|}^2}} \right)}}} \right)$ and changing the variables ${t_1} = u\cos \varphi $ and ${t_2} = u\sin \varphi $, \eqref{eq10} is rewritten as:

\begin{align}
{f_{{Z_R},{Z_I}}}\left( {{z_R},{z_I}} \right) = A\int\limits_0^\infty   &    {u{{\left( {\frac{4}{{\sigma _X^2\sigma _Y^2{{\left( {1 - {{\left| \mu  \right|}^2}} \right)}^2}}} + {u^2}} \right)}^{ - L}}}      \cdot \int\limits_0^{2\pi } {{{\rm{e}}^{ju\left( {{z_R}\cos \varphi  + {z_I}\sin \varphi } \right)}}d\varphi } \ du \ .
\label{eq11}
\end{align}

Applying the fact that $\int_0^{2\pi } {{{\rm{e}}^{x\cos \varphi  + y\sin \varphi }}d\varphi }  = 2\pi {{\rm{\mathbb{I}}}_0}\left( {\sqrt {{x^2} + {y^2}} } \right)$~\cite[eq.~3.338-4]{GradshteynBook07} and ${{\rm{\mathbb{I}}}_\alpha }\left( x \right) = {j^{ - \alpha }}{{\rm{\mathbb{J}}}_\alpha }\left( {jx} \right)$, \eqref{eq11} can be rewritten as:

\begin{equation}
{f_{{Z_R},{Z_I}}}\left( {{z_R},{z_I}} \right) = 2\pi \cdot A \cdot \int\limits_0^\infty  {\frac{{u \cdot {{\rm{\mathbb{J}}}_0}\left( {u\sqrt {z_R^2 + z_I^2} } \right)}}{{{{\left( {{B^2} + {u^2}} \right)}^L}}}du} 
\label{eq12}
\end{equation}

\noindent where $B = \frac{2}{{{\sigma _X}{\sigma _Y}\left( {1 - {{\left| \mu  \right|}^2}} \right)}}$. The integral in~\eqref{eq12} is of Hankel-Nicholson type~\cite{LukeBook62} and due to the fact that $\mathbb{K}_\alpha (x) = \mathbb{K}_{-\alpha} (x)$, \eqref{eq12} can be derived as follows:

\begin{equation}
{f_{{Z_R},{Z_I}}}\left( {{z_R},{z_I}} \right) = \frac{{2\pi A{{\left( {\sqrt {z_R^2 + z_I^2} } \right)}^{L - 1}}}}{{\Gamma \left( L \right) \cdot {{\left( {2B} \right)}^{L - 1}}}}{{\mathbb{K}}_{L - 1}}\left( {\frac{{\sqrt {z_R^2 + z_I^2} }}{B}} \right) 
\label{eq13}
\end{equation}

Substituting the expressions of $A$ and $B$ into \eqref{eq13}, we achieve the full formula of the joint PDF \eqref{eq14}, which is different compared to~\cite[eq.~(11)]{ProakisTCT68}. It can be seen that when $\mu = 0$, in our formula the joint PDF is still a function of the variances of RVs $X$ and $Y$.

\begin{figure*}[!htbp] 
    \begin{align}
	  {f_{{Z_R},{Z_I}}}\left( {{z_R},{z_I}} \right) = & \ \frac{{2{{\left( {z_R^2 + z_I^2} \right)}^{\frac{{L - 1}}{2}}}}}{{\pi  \cdot \Gamma \left( L \right) \cdot {{\left( {{\sigma _X}{\sigma _Y}} \right)}^{L + 1}}\left( {1 - {{\left| \mu  \right|}^2}} \right)}} \cdot {{\rm{exp}} \left( {\frac{{2\left| \mu  \right|}}{{{\sigma _X}{\sigma _Y}\left( {1 - {{\left| \mu  \right|}^2}} \right)}}\left( {{z_R}\cos \varepsilon  + {z_I}\sin \varepsilon } \right)}  \right)  }   \nonumber   \\
	  &   \qquad  \qquad  \qquad  \qquad  \qquad  \qquad  \qquad  \  \cdot {{\mathbb{K}}_{L - 1}}\left( {\frac{{2\sqrt {z_R^2 + z_I^2} }}{{{\sigma _X}{\sigma _Y}\left( {1 - {{\left| \mu  \right|}^2}} \right)}}} \right)
       \label{eq14}
    \end{align}
\end{figure*}

\subsection{Step 3} 

We obtain the polar coordinate form of the joint PDF of $Z_R$ and $Z_I$ by changing the variables in \eqref{eq14} as ${z_R} = r\cos \theta $ and ${z_I} = r\sin \theta $. Taking into account the Jacobian matrix determinant of $r$, we achieve the joint PDF of the amplitude $R$ and the phase $\Theta$ \eqref{eq15}. Again we can observe additional terms related to the variances of $X$ and $Y$ compared to the one derived in~\cite[eq.~(12)]{ProakisTCT68}.

\begin{figure*}[!t]
    \begin{equation}
	  {f_{R,\Theta }}\left( {r,\theta } \right) = \frac{{2 \cdot {r^L}}}{{\pi  \cdot \Gamma \left( L \right) \cdot {{\left( {{\sigma _X}{\sigma _Y}} \right)}^{L + 1}}\left( {1 - {{\left| \mu  \right|}^2}} \right)}} \cdot {{\rm{e}}^{\frac{{2\left| \mu  \right|}}{{{\sigma _X}{\sigma _Y}\left( {1 - {{\left| \mu  \right|}^2}} \right)}}r\cos \left( {\theta  - \varepsilon } \right)}}{{\mathbb{K}}_{L - 1}}\left( {\frac{2}{{{\sigma _X}{\sigma _Y}\left( {1 - {{\left| \mu  \right|}^2}} \right)}}r} \right)
       \label{eq15}
    \end{equation}
\end{figure*}

\subsection{Step 4A} 

The derivation of the PDF of $R$ is straightforward by integrating the joint PDF over the RV $\Theta$. Applying again \cite[eq.~(3.338-4)]{GradshteynBook07}, we achieve the PDF of $R$ \eqref{eq16}. Note that, this PDF can be readily applied to the performance evaluation of the time-reversal communication system, i.e., \cite{NguyenArxiv19} by setting the correlation equal to zero. For comparison, we derive the subsequent erroneous PDF of $R$ \eqref{eq16a}, $f_{\overline{R}} (r)$, based on \cite[eq.~(12)]{ProakisTCT68}:

\begin{figure*}[!t]
    \begin{equation}
	  {f_R}\left( r \right) = \frac{{4 \cdot {r^L}}}{{ \Gamma \left( L \right) \cdot {{\left( {{\sigma _X}{\sigma _Y}} \right)}^{L + 1}}\left( {1 - {{\left| \mu  \right|}^2}} \right)}} \cdot {{\mathbb{I}}_0}\left( {\frac{{2\left| \mu  \right|}}{{{\sigma _X}{\sigma _Y}\left( {1 - {{\left| \mu  \right|}^2}} \right)}}r} \right) \cdot {{\mathbb{K}}_{L - 1}}\left( {\frac{2}{{{\sigma _X}{\sigma _Y}\left( {1 - {{\left| \mu  \right|}^2}} \right)}}r} \right)
       \label{eq16}
    \end{equation}
\end{figure*}

\begin{equation}
{f_{\overline R }}\left( r \right) = \frac{{{{\left( {1 - {{\left| \mu  \right|}^2}} \right)}^L} \cdot {r^L}}}{{{2^{L - 1}} \cdot \left( {L - 1} \right)!}} \cdot {\mathbb{I}_0}\left( {\left| \mu  \right|r} \right) \cdot {\mathbb{K}_{L - 1}}\left( r \right).
\label{eq16a}
\end{equation}

\subsection{Step 4B}

By integrating the joint PDF over the RV $R$, we can achieve the PDF of the phase $\Theta$. However, the solution is often expressed in terms of the Gaussian hypergeometric series, which makes it difficult to further calculate the error probability of the \emph{M}-PSK system. The author in~\cite{ProakisTCT68} proposed an innovative way to evaluate the integral and the result is expressed in terms of the \emph{L}-th partial derivatives of the trigonometric and inversed trigonometric functions. Unfortunately, the detailed derivation is again omitted. We first re-derive the exact solution~\cite[eq.~(13)]{ProakisTCT68}. Then, we propose an alternative approximation of the mentioned integral, in which the modified Bessel function is represented by a series of elementary functions. The approximated result turns out to compose of only trigonometric functions, which can be useful to further manipulations.

\begin{itemize}
\item First method: exact solution
\end{itemize}

Changing the variable, i.e., $t = \frac{2}{{{\sigma _X}{\sigma _Y}\left( {1 - {{\left| \mu  \right|}^2}} \right)}}r$ and integrating the joint PDF ${f_{R,\Theta }}\left( {r,\theta } \right)$ over the RV $R$ yields the following PDF of the phase $\Theta$:

\begin{equation}
{f_\Theta }\left( \theta  \right) = \frac{{{{\left( {1 - {{\left| \mu  \right|}^2}} \right)}^L}}}{{\pi  \cdot {2^L} \cdot \Gamma \left( L \right)}}\int\limits_0^\infty  {{t^L} \cdot {{\rm{e}}^{\left| \mu  \right|\cos \left( {\theta  - \varepsilon } \right) \cdot t}}{{\mathbb{K}}_{L - 1}}\left( t \right)dt} 
\label{eq17}
\end{equation}

Surprisingly, \eqref{eq17} is equal to the result obtained by integrating the joint PDF \cite[eq.~(12)]{ProakisTCT68} over the variable $r$. Therefore, the subsequent results in~\cite{ProakisTCT68} can be correct. But it is not the case for the PDF \eqref{eq16} of the amplitude $R$ if used. 

In order to further compute ${f_\Theta }\left( \theta  \right)$, we first apply the integral representation of $\mathbb{K}_{L-1} (t)$~\cite[eq.~(10.32.10)]{LinkBessel} so that \eqref{eq17} can be rewritten as:

\begin{equation}
{f_\Theta }\left( \theta  \right) = \frac{{{{\left( {1 - {{\left| \mu  \right|}^2}} \right)}^L}}}{{\pi  \cdot {2^L} \cdot \Gamma \left( L \right)}}\int\limits_0^\infty  {\frac{1}{t} \cdot {{\rm{e}}^{D \cdot t}}\int\limits_0^\infty  {{{\left( {\frac{{{t^2}}}{x}} \right)}^L} \cdot {{\rm{e}}^{ - x - \frac{{{t^2}}}{{4x}}}}dx} dt} 
\label{eq18}
\end{equation}

\noindent where $D = \left| \mu  \right|\cos \left( {\theta  - \varepsilon } \right)$. We then consider the following function:

\begin{equation}
g\left( {x,c} \right) = \frac{{{t^2}}}{x} \cdot {{\rm{e}}^{ - x - \frac{{{t^2}}}{{4x}} \cdot c}} 
\label{eq19}
\end{equation}

It is obvious that the integral of $g\left( {x,c} \right)$ w.r.t $x$ converges for $x \in \left[ {0,\infty } \right)$ and $c \ge 1$. Therefore we can apply a so-called integration under the integral sign technique. Due to the fact that $ \frac{{{\partial ^{L - 1}}g\left( {x,c} \right)}}{{\partial {c^{L - 1}}}} = \frac{{{{\left( { - 1} \right)}^{L - 1}}}}{{{4^{L - 1}}}} \cdot {\left( {\frac{{{t^2}}}{x}} \right)^L} \cdot {{\rm{e}}^{ - x - \frac{{{t^2}}}{{4x}}}} $, the second integral in \eqref{eq18} is represented as follows:

\begin{equation}
\int\limits_0^\infty  {{{\left( {\frac{{{t^2}}}{x}} \right)}^L} \cdot {{\rm{e}}^{ - x - \frac{{{t^2}}}{{4x}}}}dx}  = {\left( { - 1} \right)^{L - 1}} \cdot {2^{2L - 2}}{\left. {\frac{{{\partial ^{L - 1}}g\left( {x,c} \right)}}{{\partial {c^{L - 1}}}}} \right|_{c = 1}}
\label{eq20}
\end{equation}

\noindent in which $ {\left. {\frac{{{\partial ^{L - 1}}g\left( {x,c} \right)}}{{\partial {c^{L - 1}}}}} \right|_{c = 1}} $ denotes the $L$-th partial derivative of the function $g(x,c)$ evaluated at $c=1$. Substituting \eqref{eq20} into \eqref{eq18} and after some manipulations, \eqref{eq18} can be rewritten as:

\begin{align}
{f_\Theta }\left( \theta  \right) = \frac{{{{\left( {1 - {{\left| \mu  \right|}^2}} \right)}^L}{{\left( { - 1} \right)}^{L - 1}}}}{{2\pi  \cdot \Gamma \left( L \right)}} \cdot \frac{{{\partial ^{L - 1}}}}{{\partial {c^{L - 1}}}}{\left. {h\left( c \right)} \right|_{c = 1}}
\label{eq21}
\end{align}

\noindent where $h\left( c \right) = \int\limits_0^\infty  {{{\rm{e}}^{D \cdot t}} \cdot t \cdot {{\mathbb{K}}_0}\left( {t\sqrt c } \right)dt} $. To derive $h(c)$, we use another representation of ${{\mathbb{K}}_0}\left( {t\sqrt c } \right) = \int\limits_1^\infty  {\frac{{{{\rm{e}}^{\sqrt c  \cdot t \cdot x}}}}{{\sqrt {{x^2} - 1} }}dx} $ \cite[eq.~(10.32.8)]{LinkBessel} and apply the so-called Euler substitution, i.e., $ \sqrt {{x^2} - 1}  = \left( {x + 1} \right)u $, we can achieve the following expression:

\begin{equation}
h\left( c \right) = \int\limits_0^1 {\frac{{2 - 2u^2}}{{{{\left[ {\left( {\sqrt c  - D} \right) + \left( {\sqrt c  + D} \right){u^2}} \right]}^2}}}du} 
\label{eq22}
\end{equation}

The above integral is a trivial problem. Although the derivation is algebraically complex, it is straightforward to obtain the following result:

\begin{equation}
h\left( c \right) = \frac{1}{{c - {D^2}}} + \frac{D}{{{{\left( {c - {D^2}} \right)}^{3/2}}}} \cdot \arccos \left( { - \frac{D}{{\sqrt c }}} \right)
\label{eq23}
\end{equation}

Using the definition of $D$ and substituting \eqref{eq23} into \eqref{eq21}, we obtain the PDF of the phase $\Theta$ \eqref{eq24}, which is similar as the one in~\cite[eq.~(13)]{ProakisTCT68}.

\begin{figure*}[!htbp] 
    \begin{align}
    {f_\Theta }  \left( \theta  \right)   &   = \frac{{{{\left( {1 - {{\left| \mu  \right|}^2}} \right)}^L}{{\left( { - 1} \right)}^{L - 1}}}}{{2\pi  \cdot \Gamma \left( L \right)}}  \nonumber   \\
    &  \cdot \frac{{{\partial ^{L - 1}}}}{{\partial {c^{L - 1}}}}{\left. {\left( {\frac{1}{{c - {{\left| \mu  \right|}^2}{{\cos }^2}\left( {\theta  - \varepsilon } \right)}} + \frac{{\left| \mu  \right|\cos \left( {\theta  - \varepsilon } \right)}}{{{{\left( {c - {{\left| \mu  \right|}^2}{{\cos }^2}\left( {\theta  - \varepsilon } \right)} \right)}^{3/2}}}}\arccos \left( { - \frac{{\left| \mu  \right|\cos \left( {\theta  - \varepsilon } \right)}}{{\sqrt c }}} \right)} \right)} \right|_{c = 1}}
       \label{eq24}
    \end{align}
\end{figure*}

\begin{itemize}
\item Second method: approximated solution for $L \ge 2$
\end{itemize}

We apply the elementary functions-based series representation of the modified Bessel function of the second kind proposed in~\cite[eq.~(11)]{NguyenArxiv19} as follows:

\begin{equation}
{\mathbb{K}_L}\left( {x} \right) \approx \sum\limits_{q = 0}^T {\left( {\sum\limits_{l = q}^T {\Lambda \left( {L,l,q} \right)} } \right)}  \cdot {{\rm{e}}^{ - x}} \cdot  {x} ^{q - L} ,
\label{eq25}
\end{equation}

\noindent where the number of expansion terms are limitted to $T$, ${\Lambda \left( {L,l,q} \right)}$ is a coefficient calculated by:

\begin{equation}
\Lambda \left( {L,l,q} \right) = \frac{{{{\left( { - 1} \right)}^q}\sqrt \pi   \cdot \Gamma \left( {2L} \right) \cdot \Gamma \left( {\frac{1}{2} + l - L} \right) \cdot \mathbb{L}\left( {l,q} \right)}}{{{2^{L - q}} \cdot \Gamma \left( {\frac{1}{2} - L} \right) \cdot \Gamma \left( {\frac{1}{2} + l + L} \right) \cdot l!}}
\label{eq26}
\end{equation}

\noindent and $\mathbb{L}\left( {l,q} \right) = \left( {\begin{array}{*{20}{c}}{l - 1}\\{q - 1}\end{array}} \right)\frac{{l!}}{{q!}}$ for $\forall l, q > 0$ is the Lah number~\cite{NguyenArxiv19} with the conventions $\mathbb{L} (0,0) = 1$, $\mathbb{L} (l,0) = 0$, $\mathbb{L} (l,1) = l!$ for $\forall l > 0$. Since the order of the modified Bessel function of the second kind in \eqref{eq17} $L-1$ is an integer number bigger than $1$, the series representation of $\mathbb{K}_{L-1} (t)$ converges~\cite{NguyenArxiv19}. 

Although the derivation is algebraically complicated, it is straightforward to obtain the tractable closed-form approximation of the PDF of $\Theta$ \eqref{eq27}. It can be observed that the formula of ${f_\Theta }\left( \theta  \right)$ is convenient to derive the error probability of phase-modulated signals, as the calculation of the error probability is equivalent to compute the integrals of the form $\int_{{\theta _1}}^{{\theta _2}} {\frac{{d\theta }}{{{{\left( {a + b\cos \theta } \right)}^n}}}} $, whose results exist in the literature, i.e., \cite[eqs. (2.554)]{GradshteynBook07}, or it can be easily numerically evaluated. The particular cases of error probability calculations are out of the scope of this letter.

\begin{figure*}[!htbp] 
    \begin{equation}
	    {f_\Theta }\left( \theta  \right) \approx \frac{{{{\left( {1 - {{\left| \mu  \right|}^2}} \right)}^L}}}{{\pi  \cdot {2^L} \cdot \Gamma \left( L \right)}}\sum\limits_{q = 0}^T {\sum\limits_{l = q}^T {\frac{{\Lambda \left( {L - 1,l,q} \right) \cdot \Gamma \left( q+2 \right)}}{{{{\left( {1 - \left| \mu  \right|\cos \left( {\theta  - \varepsilon } \right)} \right)}^{q + 2}}}}} } ,{\rm{  \quad  if  \quad  }}L \ge 2
       \label{eq27}
    \end{equation}
\end{figure*}

\section{Numerical Results}
\label{sec2}

For the sake of simplicity, we set variances $\sigma_{X} = 0.7$, $\sigma_{Y} = 1.5$ and the correlation $\mu = 0.5 \cdot \exp \left( {j\pi /6} \right)$ in all simulations. The simulated histograms are plotted after 100~000 realizations.

\begin{figure*}[!htbp]
\centering
\includegraphics[width=0.9\textwidth]{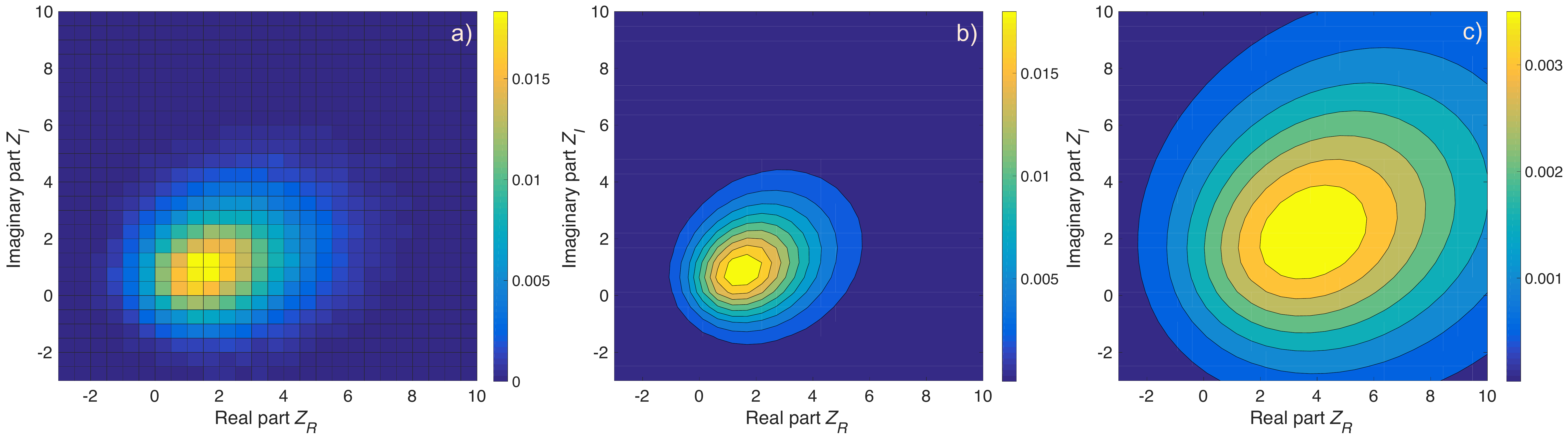} 
\caption{Normalized 2-D histogram of the RV $ Z $ (constructed by the sum of five products ($L=5$) of 2 complex Gaussian zero-mean RVs, variances $\sigma_{X} = 0.7$, $\sigma_{Y} = 1.5$ and correlation $\mu = 0.5 \cdot \exp \left( {j\pi /6} \right)$). a) Simulated histogram; b) Analytic histogram using \eqref{eq14}; c) Analytic histogram using \cite[eq.~(11)]{ProakisTCT68}.}
\label{fig1}
\end{figure*}

We first numerically validate the derived joint PDF \eqref{eq14}. Fig.~\ref{fig1} presents the normalized two-dimensional histogram of the RV $Z$, when $L = 5$. The simulated histogram is illustrated in Fig.~1(a). The analytic histograms applying \eqref{eq14} and \cite[eq.~(11)]{ProakisTCT68} are plotted in Fig.~1(b) and Fig.~1(c), respectively. It can be observed that the histogram built based on our derivation looks similar to the simulated histogram while the one built based on the derivation in~\cite{ProakisTCT68} is much wider.

\begin{figure*}[!htbp]
\centering
\includegraphics[width=1.0\textwidth]{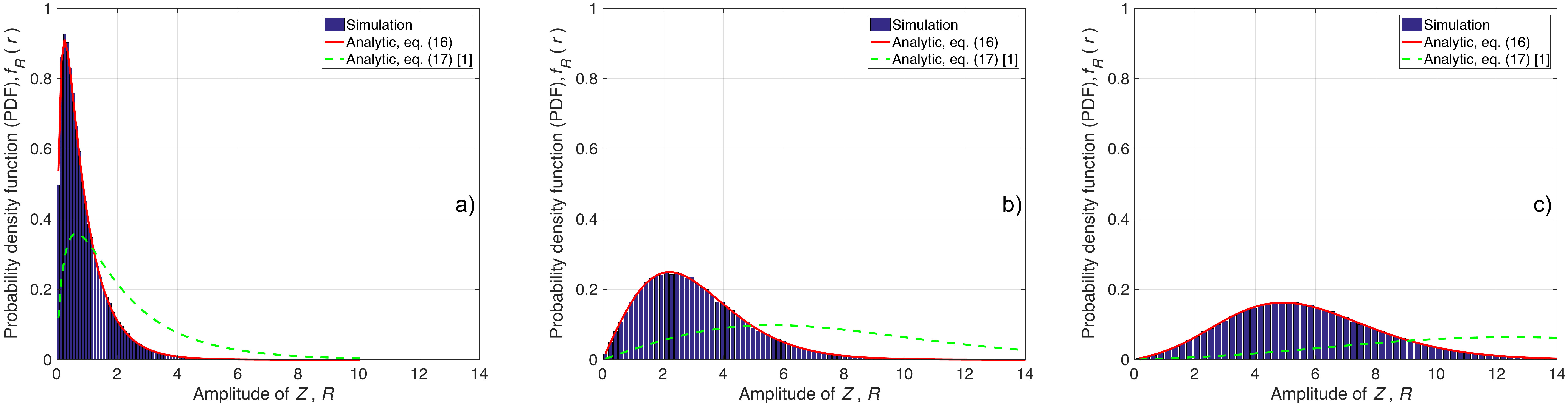} 
\caption{Histogram of the amplitude of RV $ Z $ and their associated analytical PDFs $f_R (r)$, when variances $\sigma_{X} = 0.7$, $\sigma_{Y} = 1.5$ and correlation $\mu = 0.5 \cdot \exp \left( {j\pi /6} \right)$. a) $L = 1$; b) $L = 5$; c) $L = 10$.}
\label{fig2}
\end{figure*}

\begin{figure*}[!htbp]
\centering
\includegraphics[width=1.0\textwidth]{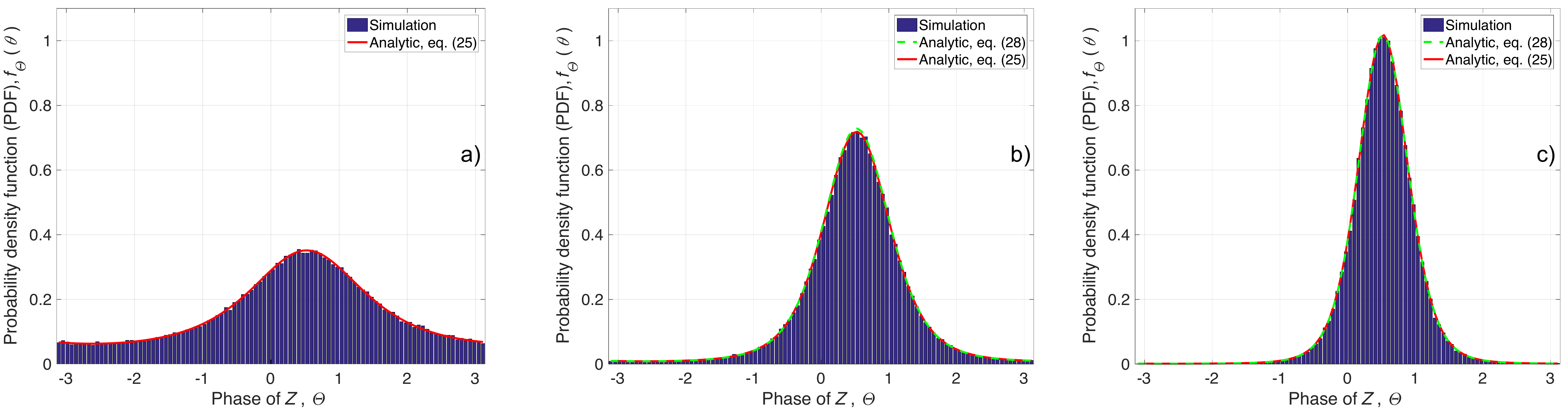} 
\caption{Histogram of the phase of RV $ Z $ and their associated analytical PDFs $f_\Theta (\theta)$, when variances $\sigma_{X} = 0.7$, $\sigma_{Y} = 1.5$ and correlation $\mu = 0.5 \cdot \exp \left( {j\pi /6} \right)$. a) $L = 1$; b) $L = 5$; c) $L = 10$.}
\label{fig3}
\end{figure*}

Since the derivation of \eqref{eq15} was with a simple manipulation from \eqref{eq14}, we directly validate derived PDFs of the amplitude of $Z$ \eqref{eq16} and of the phase of $Z$ \eqref{eq24} and \eqref{eq27}. To reduce the complexity, the number of expansion terms in \eqref{eq27} is set to $L$, i.e., $T=L$. In what follows, we consider three cases, in which (a) $L = 1$; (b) $L = 5$ and (c) $L = 10$.

Figure~\ref{fig2} presents the histograms of the amplitude of $Z$ and their associated analytical PDFs $f_R (r)$ \eqref{eq16} derived in this letter and $f_{\overline{R}} (r)$ \eqref{eq16a} derived based on result in~\cite{ProakisTCT68}. It can be seen that our proposed PDF fits better the actual PDF, confirming the correctness of our derivation.

Figure~\ref{fig3} plots the histograms of the phase of $Z$ and their associated analytical PDFs $f_{\Theta} (\theta)$ \eqref{eq24}, \eqref{eq27}. Both PDF envelopes match each other and fit well the simulated histograms. Although the results for $L=2$ and $L=3$ are omitted, it is worth noticing that our proposed approximated PDF provides a good approximation whenever $L \ge 4$, which frequently happens in the large bandwidth and densely connected systems. Furthermore, our proposed solution of the PDF of the phase of $Z$ is better suited to the further manipulations and does not require extra partial derivative operator. As the rest of~\cite{ProakisTCT68} was based on the PDF of the phase of $Z$~\cite[eq.~(13)]{ProakisTCT68}, which is correct, the results of this paper is still valid.

\section*{Acknowledgment}

The authors would like to thank the financial support of the Copine-IoT Innoviris project, the Icity.Brussels project and the FEDER/EFRO grant.

\end{document}